%% file: 0_main.tex
\documentclass[runningheads]{llncs}
\usepackage{cite}
\usepackage{graphicx}
\usepackage{comment}
\graphicspath{ {./images/} }

\def \auc{\textsc{AUC}}
\def \tool{\textsc{EduPal}}

\begin{document}
\title{\tool \space leaves no professor behind: Supporting faculty via a peer-powered recommender system}
\titlerunning{Leave No Professor Behind}

\author{Nourhan Sakr \inst{1}\orcidID{0000-0002-6130-9795} \and
Aya Salama\inst{1}\orcidID{0000-0001-7111-7677}\and
Nadeen Tameesh\inst{1}\orcidID{0000-0002-1506-1004} \and
Gihan Osman\inst{1}\orcidID{0000-0001-8830-3409}}
\authorrunning{Sakr et al.}
\institute{The American University in Cairo, Cairo, Egypt\\
\email{n.sakr@columbia.edu,}
\email{\{aya{\_}salama,nadeentameesh14,gosman\}@aucegypt.edu} 
}

\maketitle              % typeset the header of the contribution
\begin{abstract} 
The swift transitions in higher education after the COVID-19 outbreak identified a gap in the pedagogical support available to faculty. We propose a smart, knowledge-based chatbot that addresses issues of knowledge distillation and provides faculty with personalized recommendations. Our collaborative system crowdsources useful pedagogical practices and continuously filters recommendations based on theory and user feedback, thus enhancing the experiences of subsequent peers. We build a prototype for our local STEM faculty as a proof concept and receive favorable feedback that encourages us to extend our development and outreach, especially to underresourced faculty. 

\end{abstract}

\section{Background and Related Work}
\input{01_Motivation}

\section{Data-Driven Modeling}
\input{02_Best_practices}

\section{Recommender System Design}
\input{03_recommender_system}

\section{Prototype Evaluation and Conclusion}
\label{conc}
\input{5_evaluation}

\bibliographystyle{splncs04}
\bibliography{references}

\end{document}

%% file: 01_Motivation.tex
The COVID-19 lock-down forced many higher education institutions globally to continue instruction via online modalities at an unprecedented pace and scale \cite{hodges2020difference, czerniewicz_2020}. With many faculty scantily trained in teaching strategies or with little support on best online practices  \cite{cutri2020critical,hodges2020difference, xie2021instructional}, instruction was maintained at the cost of education quality, equity and sound pedagogy \cite{crawford2020, motala2020search,czerniewicz_2020}. Online education requires deliberate design and development \cite{hodges2020difference, means2014learning,czerniewicz_2020,xie2021instructional}, yet, the pandemic forced the adoption of \textit{emergency remote learning}, regardless of any obstacles.

In non-emergency times, faculty in resourced institutions are often supported by instructional designers who provide personalized guidance on making sound design and technology decisions for the faculty's particular context \cite{beirne2018instructional}. However, given the sheer number of ``overnight'' transitions, individualized help became rather challenging \cite{hodges2020difference}.  %Moreover,
The pandemic revealed the lacking capacities for support and infrastructure in institutions \cite{crawford2020,kimmons_veletsianos_vanleeuwen_2020,leung_sharma_2020,wu_wu_zhaohui_2020}, thereby questioning readiness for the digital era. Looking further into %underprivileged contexts of 
under-resourced institutions, general capacity building and high-quality instructional guidance are considered a luxury.

In light of this extreme global test, we identify a gap in the pedagogical support available to educators. Social media %, such as the \textit{Pandemic Pedagogy} group on Facebook and free 
 and online webinars %, such as the ones provided by \textit{Harvard Education},
attempted closing this gap by providing platforms for sharing experiences and sound tips online. However, we see three issues with such channels: They are (1) less personalized, (2) suffer from information overflow and (3) are not guaranteed to continue after the pandemic. These issues make it hard for some faculty to find relevant resources or apply what they find to their personalized contexts. Based on our survey ($n=103$), 86\% believed that being able to readily access relevant experiences shared by peers would be beneficial to them, even in the long run.

Within this framework, we propose, \tool, a virtual educational consultant. Our user-centric design provides a crowd-sourcing platform augmented with collaborative filtration to automate experience sharing and knowledge distillation. We wrap these within a recommendation system that provides personalized, context-aware guidance on best-fit pedagogical practices, as supported by research theory and faculty practice. As a proof of concept, \tool\space was customized for our local STEM community at the American University in Cairo (\auc), classifies as an M1 university \footnote{According to the Carnegie Classification of Institutions of Higher Education}. In this paper, we present our pilot's data collection methodology, system design and show positive user feedback declaring the system as promising for extension and generalization.

%% file: 02_Best_practices.tex
Our data-driven design builds on a taxonomy %\cite{ourpaper} 
that is the product of an elaborate data collection process, outlined briefly in this section. Despite our focus on supporting STEM faculty at \auc \space (for the pilot study), we consider various populations to build our data. %As such, we compile narratives from global faculty and students on social media platforms, interview faculty at \auc \space and education experts around the world.
We apply maximum variation sampling in recruiting participants and conclude any stage when the research team agrees on information saturation. Our findings heavily rely on qualitative analysis. Our secondary research builds on education and psychology literature, as well as social media narratives from all stakeholders, i.e. faculty, students and instructional designers. 

\paragraph{Community Feedback.} The first stage distills knowledge from 50 hours of semi-structured interviewing of faculty at \auc, spanning all schools. Faculty reflected on their teaching challenges, need for pedagogical support, and practices most effective for their specific class types% and more likely to be adopted after the pandemic
. Results were augmented with secondary research to determine sets of: 1)features that identify profiles of instructors/classes and 2)pedagogical practices and technology tools that best fit each profile.

\paragraph{Filtration and Validation for STEM.} The second stage refines the features and recommendations to those most applicable to STEM courses via a semi-open survey aimed at STEM faculty at \auc \space ($n=100$).% Given our interest in supporting underprivileged communities, we also filtered our recommendations for ones that are more economic in application.Our refined taxonomy \cite{ourpaper} reflects the community's thoughts on best practices for delivering content, engaging students and conducting assessments.
We, then, ran two seminars for STEM faculty at \auc \space ($n=50$) and two seminars for instructional designers ($n=14$). Our recommendations were presented for discussions regarding their viability and feasibility. This process seeds our recommendations bank.

\paragraph{Learning from Experts.} Finally, we conducted semi-structured interviews with global educational consultants ($n=9$, 60 mins each). The interviews simulated a pedagogical consultation followed by a discussion on the usual process. The goal was to observe and model the thought process of an instructional designer during a consultation and identify the features they look for to formulate a recommendation. This model sets the question flow in \tool.

%% file: 03_recommender_system.tex
\begin{figure}[t]
    \centering
    \vspace{-15pt}
    \includegraphics[width=\linewidth]{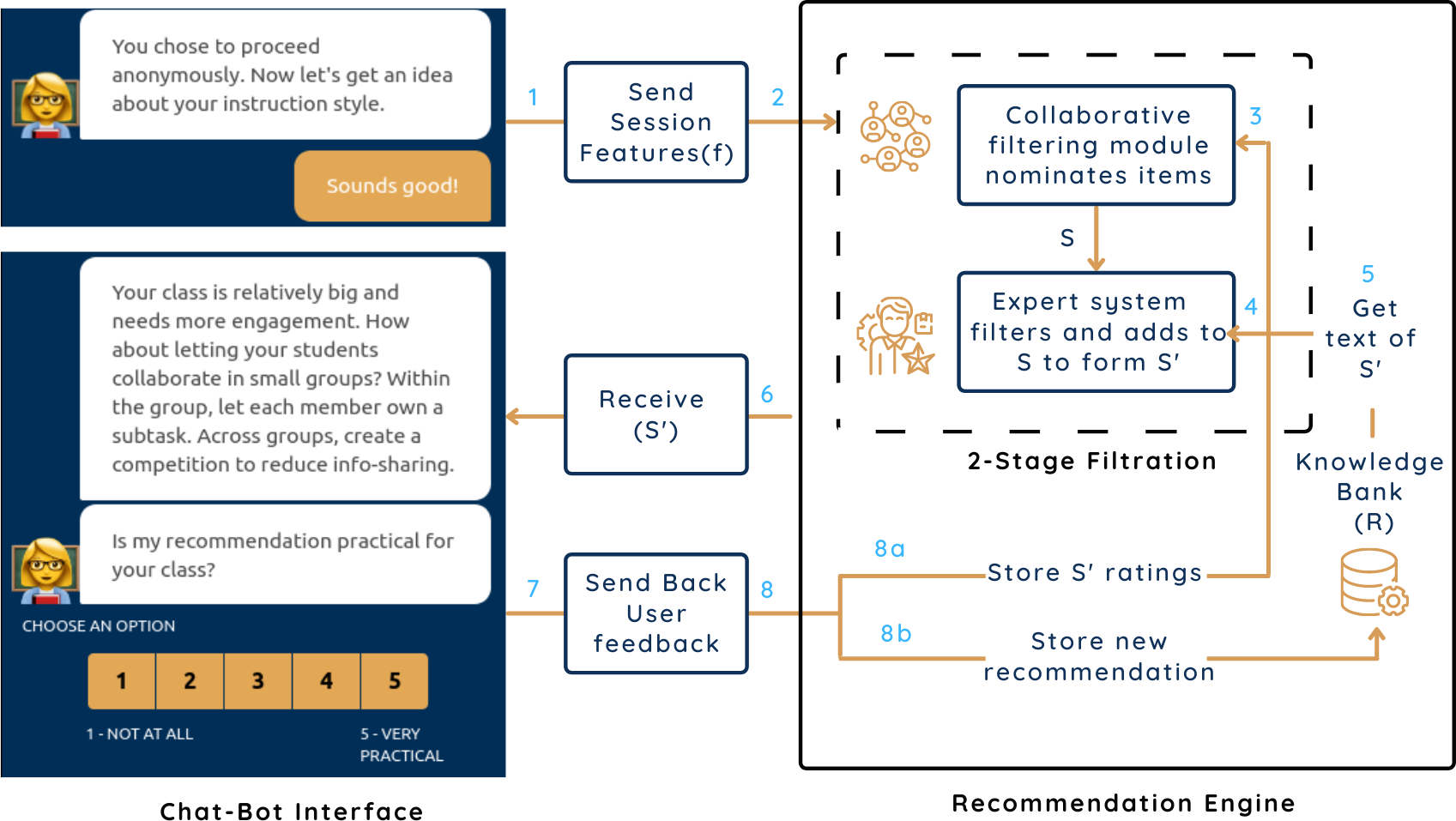}
    \caption{System components and flow.
    (1)\tool \space collects session features, \textit{f} (2)\textit{f} are sent to the recommendation engine (3)Collaborative filtering module selects \textit{S}, from the knowledge bank(\textit{R}) based on \textit{f} (4)Expert system refines \textit{S}, forming \textit{S'} (5)\textit{S'} text retrieved from \textit{R} (6)Items in \textit{S'} presented sequentially to the user (7)Ratings are assigned to S' by the user (8a) Ratings and (8b) new recommendations are stored}
    \label{fig:Edu_bot_system}
\end{figure} 

\tool \space is an instance of knowledge-based recommender systems \cite{ricci2011introduction, Garcia-Martinez2013, klavsnja2015recommender, burke2000knowledge, mahmood2009improving, garcia2009architecture,Ramadoss2006ManagementAS}. 
Via a chatbot interface, the user (assumed to be an instructor) %builds trust and 
interacts with the system as if talking to an instructional designer. The conversational session collects \textit{session features} ($f$) about the user and their course. The design encompasses a %compound filtering approach, as will be described below. The 
two-staged pipeline that allows users to benefit from the perspectives of practicing peers and pedagogical experts, as well as rate and suggest recommendations, thus automating experience sharing. The proposed recommendations are meant to enhance the three class room interaction modes defined in \cite{moore1989three}. Communication with the server is secure and confidential. An ``anonymous'' mode is added to maintain privacy of users who do not want to share their information. Figure \ref{fig:Edu_bot_system} depicts the system components and flow.

\paragraph{Collaborative filtering based on educators feedback.}
A user-based collaborative filtering approach\cite{ricci2011introduction} is used to compile an initial set of recommendations. When a new session starts, $f$ is collected and 
%profiles most similar to the current profile are identified through similarity scoring. Currently, 
cosine similarity~\cite{cosine_sim} is used %as a distance measure. 
to identify the most similar users and extract the set of recommendations $S$, %$S\subset R$, 
top rated by those users%for the most similar users are extracted to form a set of candidate recommendations
. This approach is suitable only for recommendations that have received ratings, while unrated %by a previous system user.
recommendations are sent to the {\em expert system}, which is authorized to update the knowledge bank $R$.

\paragraph{Expert system.}

This module mimics the decision making processes of an instructional designer and addresses the cold start problem\cite{coldstart}. It can be considered as a symbolic AI system where rules are constructed based on feedback from subject matter experts \cite{symbolic}. It ensures that the recommendations are not only based on popularity but are also pedagogically sound with %balanced 
support from research and practice. %$\forall r \in R$ 
For each recommendation $r \in R$, experts identify the factors defined in $f$ that should match $r$ with a user, based on the learning sciences. Those rules are then translated into system logic. %For  in our database, rejection and acceptance criteria were identified by expert consultants (whether collaborators or interviewees) and then translated into system logic. 
where, each element $s \in S$)is either accepted as part of the final set, $S'$, or rejected. This logic is also used to add recommendations, deemed as best fit by experts, to $S'$.
\input{4_3_feedback}

%% file: 4_3_feedback.tex
\begin{comment}
    
\begin{figure}
    \centering
    \vspace{-15pt}
    \includegraphics[scale=0.3]{images/feedback.jpg}
    \caption{Feedback Interface}
    \label{fig:Feedback}
\end{figure} 
\end{comment}

\paragraph{Feedback.}Each recommendation in $S'$ is presented in a conversational format and rated by the user. The ratings are fed-back to the system to inform future selections\footnote{Future updates will verify that the user is faculty and that their recommendation is supported by research before incorporating their feedback/rating.}. Finally, the user may share other effective pedagogical practices, that are considered for extending $R$ and fully streamlining experience sharing.

%% file: 5_evaluation.tex
\auc \space faculty\footnote{80\% were STEM, 70\% were female, experience ranged from 2 to 32 years.} ($n=10$) evaluated the prototype via 30-minute usability tests. We first learned about (1) their means and frequency of seeking help with pedagogical matters and (2) their experience with chatbots. They, then used \tool \space and provided ratings on the overall quality of the experience and the received recommendations. On a Likert scale out of 5, the mean responses to those questions were  3.8 \& 3.7, respectively. Testers were also asked to share what they liked and disliked about \tool\space as well as its advantages and disadvantages over their default methods of seeking help. The majority found \tool \space beneficial and user-friendly. They highlighted fast feedback and constant availability as immediate advantages over other aid methods. The chatbot interface made their experience feel interactive, engaging and personalized.
Our recommendations were deemed of good quality but users suggested providing more specific examples for application. Lastly, \tool's anonymous mode provided a ``safe zone'' for instructors who usually avoid sharing experiences or asking for help.

Based on the favorable feedback received, we conclude that \tool \space shows a successful pilot system and is worth generalizing to address global knowledge distillation, experience sharing automation, recommendation personalization and support scalability, in addition to promoting fairness in access to resources, given that \tool \space is available 24/7 for free. We  also recognize the potential of \tool\space becoming a screening tool for educational consultations, thus augmenting the impact of instructional designers and making appointments run faster in times of high demand, e.g. during the pandemic.